# CARBON COATING OF THE SPS DIPOLE CHAMBERS

P. Costa Pinto, S. Calatroni, P. Chiggiato, P. Edwards, M. Mensi, H. Neupert, M. Taborelli, C. Yin Vallgren, CERN, Geneva, Switzerland.

## Abstract

The Electron Multipacting (EM) phenomenon is a limiting factor for the achievement of high luminosity in accelerators for positively charged particles and for the performance of RF devices. At CERN, the Super Proton Synchrotron (SPS) must be upgraded in order to feed the Large Hadron Collider (LHC) with 25 ns bunch spaced beams. At such small bunch spacing, EM may limit the performance of the SPS and consequently that of the LHC. To mitigate this phenomenon CERN is developing a carbon thin film coating with low Secondary Electron Yield (SEY) to coat the internal walls of the SPS dipoles beam pipes. This paper presents the progresses in the coating technology, the performance of the carbon coatings and the strategy for a large scale production.

## INTRODUCTION

The SPS is the last injector of the accelerator chain that feeds the LHC with particles. To fulfill high beam brilliance and the intensity required for the future High-Luminosity LHC, (HL-LHC), the SPS must be upgraded in order to be able to inject 25 ns bunch spaced beams, with $2.2 \times 10^{11}$ protons per bunch. Such short spaced beams induce EM and may lead to dynamic pressure rise, transverse emittance blow up, thermal load and beam losses, limiting the performance of the SPS and consequently that of the LHC. For a certain type of beam, the threshold to induce EM depends on the cross section of the beam pipe, the maximal SEY, $\delta_{max}$, of its internal walls and the electrical and magnetic fields applied [1]. One way to mitigate or even eliminate EM is to reduce $\delta_{max}$ below the threshold limit. For about 80% of the total length of the machine, filled with Magnetic Bending dipoles of types A and B, (MBA an MBB), the calculated $\delta_{max}$ allowed to avoid EM are 1.4 and 1.3 respectively [2]. The beam pipes of these dipoles are made of stainless steel, are not bakeable and have a $\delta_{max}$ "as received" of about 2.0. This value can go down due to beam conditioning and preliminary results indicate this may be enough to reduce the EM in the MBA dipoles to acceptable levels [3]. If this is confirmed, this type of dipoles will not need to be coated. For the MBB type, the level of EM may remain high even after beam conditioning. Further decrease of EM can be obtained by coating the internal walls of the beam pipes with a low SEY thin film.

For this purpose, CERN has been developing carbon coatings with a $\delta_{max}$ of about 1.0 which are robust against air exposure for long times. Because the geometry of the beam pipes and the way they are available for coating varies considerably, several coating techniques and setups have been tested and developed. To coat new beam pipes and then insert them in the magnets, Direct Current Cylindrical Magnetron Sputtering (DCCMS) from graphite targets has proven to give quality coatings. If all the issues concerning industrial scale production are solved, this approach is expensive and risky since it implies to disassemble / reassemble all the magnets to insert the coated beam pipe (with the exception of quadrupoles). To coat the actual beam pipes in the dipoles, without disassembling / reassembling, several techniques have been tested: Plasma Enhanced Chemical Vapor Deposition (PECVD); two versions of Direct Current Planar Magnetron Sputtering (DCPMS), one with permanent magnets and another using the dipole's own field; and Direct Current Hollow Cathode Sputtering (DCHCS). This last coating technique is considered the most adequate for a large scale production and is presently being optimized. Figure 1 shows the SEY as a function of the energy of primary electrons for stainless steel as received, after conditioning in the SPS and for carbon thin films obtained by different coating techniques used at CERN.

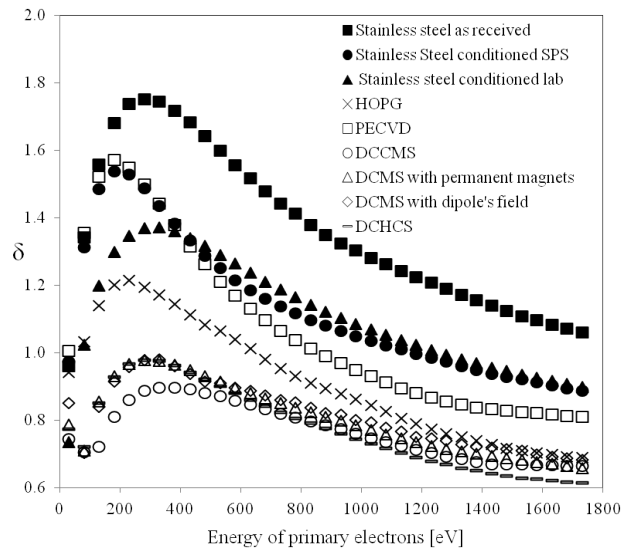

Figure 1: SEY as a function of the energy of the primary electrons for stainless steel and carbon coatings. The SEY curve for Highly Ordered Pyrolytic Graphite (HOPG) is also plotted for comparison.

The carbon thin films manufactured by the different techniques undergo different levels of testing depending on their potential to be applied in the SPS. In the laboratory: SEY measurements and EM induced by a RF standing wave using a dipole as a coaxial resonator [4]; in the SPS: Electron Cloud Monitors, (ECM), [5] and dynamic pressure rise. The last phase in the validation of the carbon thin films is the installation of two SPS cells with coated beam pipes. Four MBB type dipoles, two MBA type and one quadrupole are already installed. During the first long shutdown of the SPS in 2013, 14 other dipoles with coated chambers are planned to be installed.

## COATING TECHNIQUES AND SETUPS

Figure 2 shows the cross sections of the main beam pipes to be coated. Both MBB and MBA types have a total length of 6.5 meters and are embedded in an 18 ton dipole. Only a strip of ~60 mm in width, centred on the top and bottom needs to be coated.

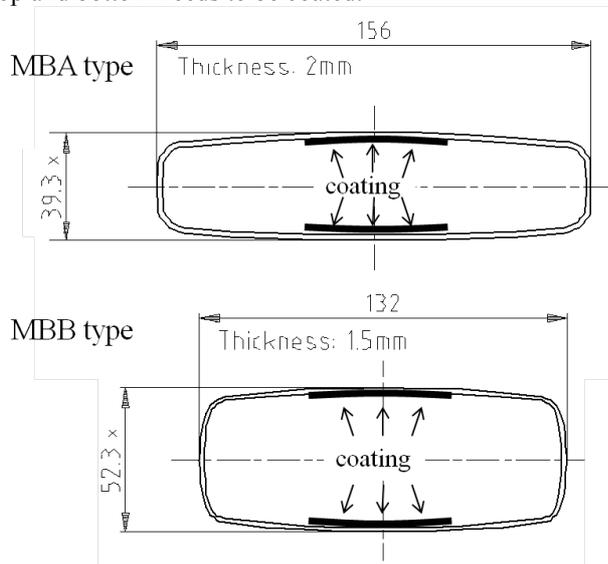

Figure 2: Cross sections of the main dipoles beampipes of the SPS.

### DCCMS

New beam pipes don't have flanges and must be housed in a vacuum chamber for the coating process. The carbon target is made of two 7.5 meter long graphite rods φ13 mm, (ashes content <400ppm), held by a stainless steel structure that also masks the sides of the beam pipes where the carbon coating is not necessary and centres the ensemble in the beam pipe (Figure 3). Two additional targets, made of Ti, Zr and V wires, allow the co-deposition of getter on the side walls. The role of the getter is to reduce the partial pressure of hydrogen in the plasma during the growth of the carbon film. A correlation between the hydrogen partial pressure and the SEY was found [6] and confirmed by experiments where hydrogen was deliberately mixed with the discharge gas during film growth. The target assembly is 7.5 m long and the insertion in the beam pipe is done vertically. The ensemble is inserted in a 8 meter long solenoid that provides the magnetic field to confine the electrons in the magnetron discharge (180 Gauss) [6]. After pumpdown and bakeout at 300$^o$C for 24 hours, the discharge gas, Ne, is injected to a pressure of about $4\times10^{-2}$ mbar and the plasma ignited. For a power of about 2 kW, (700 V, 2.8 A), the deposition rate is about 70 nm/hour for the case of MBB beampipes (100 nm/hour for the MBA). At lower pressures and power the plasma formed rings along the cathodes, resulting in a non uniform coating thickness. During deposition the temperature of the substrate rises up to 280$^o$C.

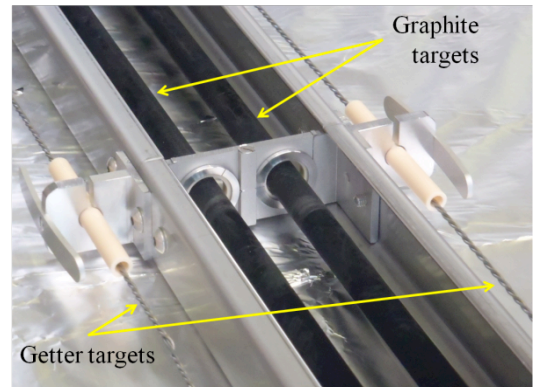

Figure 3: Cathode configuration to coat MBA chambers by DCCMS. A getter is co-deposited on the sides of the beam pipe in order to reduce the hydrogen partial pressure during deposition. Two masks screen the graphite targets.

### PECVD

When used in hollow cathode configuration, with the beampipe being the cathode, PECVD allows the deposition of thin films without the insertion of electrodes. This could simplify the coating setup, allowing to coat in the beampipes without disassembling the dipoles. Figure 4 shows a photo of the coating system.

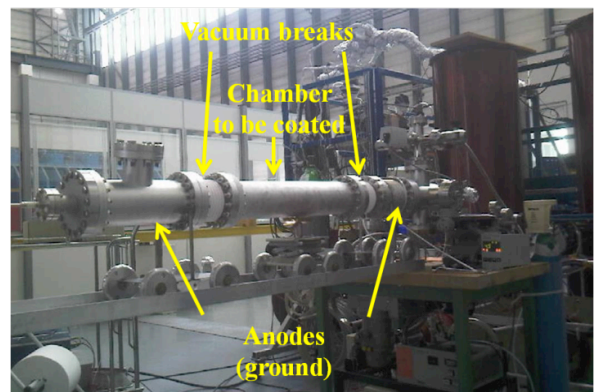

Figure 4: The system used to produce carbon coatings from acetylene by PECVD.

Acetylene, $C_2H_2$, was used as reactive gas. The pressure during deposition was about $10^{-2}$ mbar, the power ranged from 50 W to 200W and the voltages from 500V to 1800V. Deposition rates could reach 1 μm/hour.

## DCPMS with the dipole's field

Figure 5 a) shows a sketch of the electrode's setup used in this technique. The targets, made of graphite 30 mm x 10 mm rectangular bars (type R7200 from Steinemann AG), are placed parallel to the dipole's magnetic field. Stainless steel anodes, in front of each graphite bar, deform the electric field to maximize the magnetron effect (electric and magnetic fields perpendicular to each other). A glow discharge develops between the graphite and the anode and the sputtered carbon atoms are deposited on the top and bottom of the beampipe.

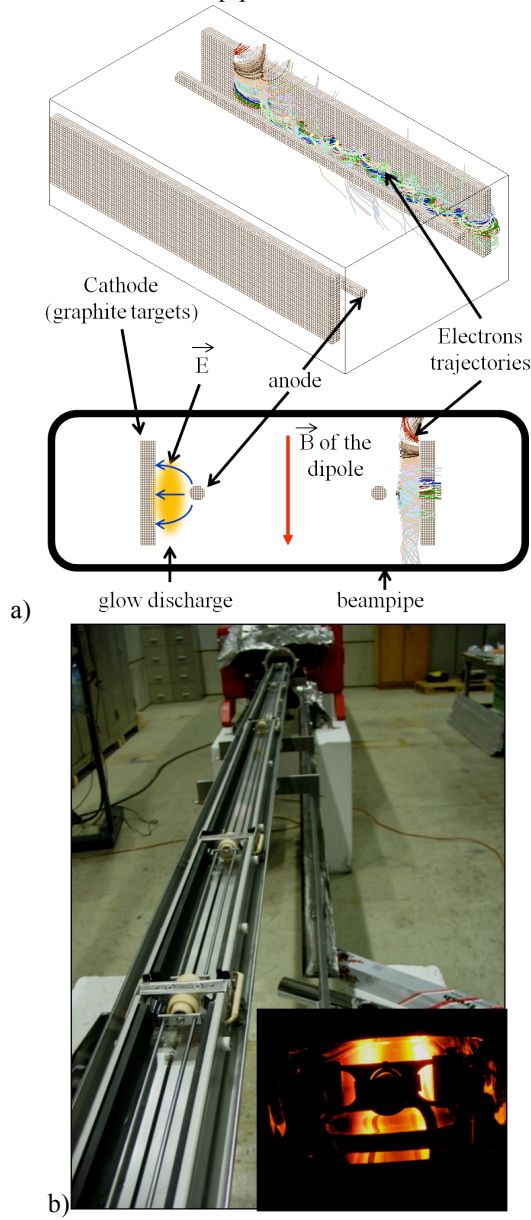

a)

b)

Figure 5: Electrodes configuration for DCPMS using the dipole's magnetic field: a) Electrons trajectories simulated in SimIon 7.0, (from Idaho National Engineering & Environmental Laboratory, USA; b) view of the 6.5 meter long electrode and of the resulting glow discharge.

The 6.5 meter long target assembly to coat a full dipole is made of units of 1 meter long connected in series (Figure 5 b)). Plasma confinement is not very efficient because the electrons are lost at the extremities of each unit (Figure 5 a)). This loss of electrons results in a gradient in the plasma density along each target unit and consequently a non uniform deposition rate. To compensate this effect the polarity of the magnetic field was swapped regularly during the coating. The discharge pressure was $9 \times 10^{-1}$ mbar (Ne), the total power 2 kW (at 900V, 2.1A) and the temperature of the substrate 140°C. Film thickness was not uniform, with ~200nm in the centre and 1100 nm at the extremities (close to the graphite targets).

## DCPMS with permanent magnets

In this setup, the magnetic field to confine the electrons is generated by an array of small magnetic circuits made of permanent magnets. Each magnetic circuit consists in a pair of magnets, an inner cylindrical magnet, (R=4 mm, h=10 mm) and a outer ring magnet, ($R_{in}$=10 mm, $R_{out}$=15 mm, h=10 mm), made from $Sm_2Co_{17}$. Figure 6 a) shows a photo of a unit 1 meter long, with an array of 32 magnetic circuits, before welding the top half of their container. Two graphite targets, made of 800 mm x 70 mm x 2mm plates (grade 2020PT from Mersen), are placed on the top and bottom of the container and isolated by 0.5 mm thick alumina spacers. Only one prototype of 1 meter long was built and tested. To keep the temperature of the magnets below their critical temperature, 150°C, the maximal power was 50 W. The bleeding gas was Ne at a pressure of $6 \times 10^{-2}$ mbar.

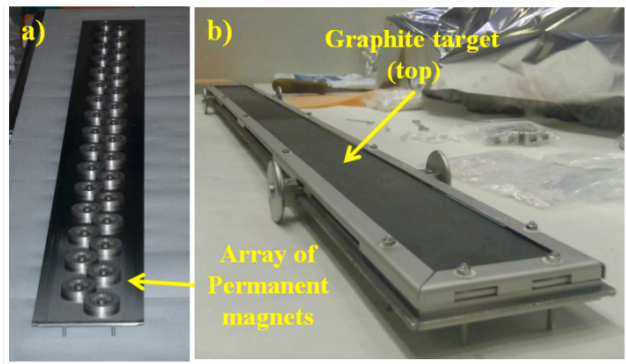

Figure 6: The magnetic field is supplied by an array of permanent magnets; a) view of the 32 magnetic circuits; b) 1 meter unit of the complete assembly.

Due to the reduced size of the magnetic circuit, the field strength decreases strongly the further away from the target surface we go. For this reason the plasmas will tend to concentrate in the places where the target is closer to the magnetic circuit resulting in an unevenly distributed sputtering.

## DCHCS

In the hollow cathode discharge, the electrons are confined by the potential of the cathode walls. This principle can be applied to the dipole's geometry by building graphite hollow rectangular cells. Electrons leaving each of the targets faces are reflected by the other ones, increasing their path inside the cell until being lost in the anode's surface (the beampipe). Though this particular geometry doesn't result in a very efficient hollow cathode it is enough to confine the plasma inside the cell. The 6.5 meter long target assembly to coat the dipoles chambers without dismounting /remounting is made of units of 1 meter electrically connected in series. To improve thickness uniformity the target assembly is continuously moved back and forth along the axis of the beampipe. The discharge power is 1800 W, (I=3 A, U=600 V), resulting in a deposition rate of 30 nm per hour and the discharge gas is Ar. Temperature during the coating stabilizes about 50$^o$C.

## RESULTS

### Coatings morphology

Images of the coatings morphology were obtained by Scanning Electron Microscopy, (SEM), in a LEO 430I instrument. From Figure 7 we can see that all the coatings are rougher than the highly ordered pyrolytic graphite. Those obtained by PECVD are less rough than the sputtered ones. The adhesion is evaluated by a simple scotch tape test and the coatings issued from the different techniques have successfully passed this test.

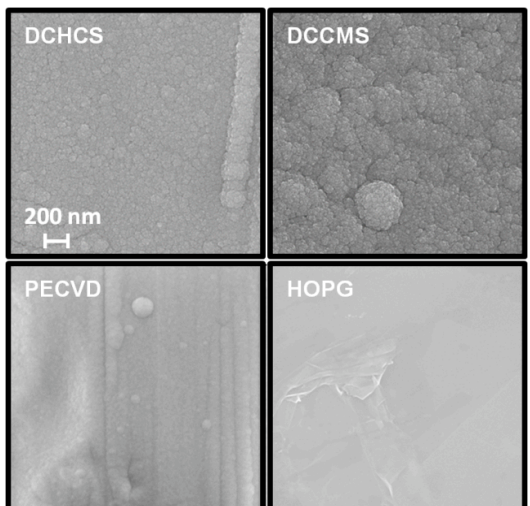

Figure 7: SEM images of the coatings produced by different techniques. HOPG is displayed for reference.

### Pump down characteristics

To check the compatibility of the coatings with vacuum, the evolution of the pressure with time during pump down was monitored. The results presented in Figure 8 were obtained for a 6.4 meter MBB beampipe before and after coating by DCHCS. The chamber was pumped through a 1 meter long bellow with 35 mm internal diameter by a turbo molecular pump and the pressure measured by a penning gauge. After 20 hours of pumping the pressure in the carbon coated chamber was about 3.5 times higher than in the stainless steel one.

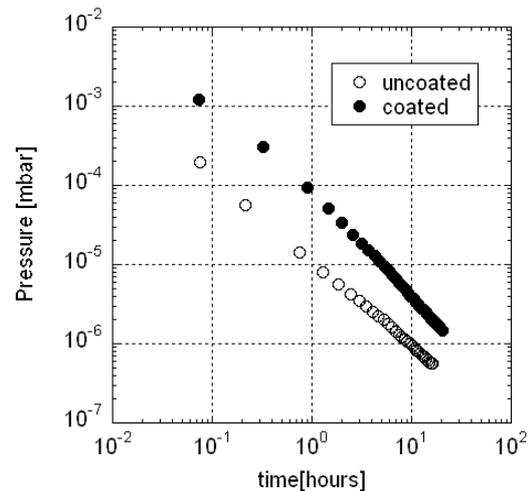

Figure 8: Pump down characteristics for a MBB beampipe before and after coating with carbon.

### Measurement of the $\delta_{max}$ in laboratory

Laboratory measurements of the SEY for coatings issued from the different techniques are plotted in Figure 1. After deposition, the coatings are vented to 1 bar of dry air for about 15 minutes and then exposed to the laboratory air. The coatings done by the different sputtering setups have all $\delta_{max}$ around 1 while PECVD ones are above 1.5. Sputtered coatings have lower SEY than HOPG graphite. Despite the wide range of parameters explored for the sputtered coatings, power density (80 W/m up to 400 W/m), voltage (400V up to 900 V), pressure (8x10$^{-3}$ mbar up to 9x10$^{-1}$ mbar), discharge gas type (Ne, Ar and Kr), substrate temperature (100$^o$C up to 350$^o$C) and hydrogen partial pressure in the discharge gas, only this last one was found to influence in the SEY. Figure 9 shows the values of $\delta_{max}$ for coatings done by DCCMS with different partial pressures of hydrogen in the discharge gas and for different coating temperatures.

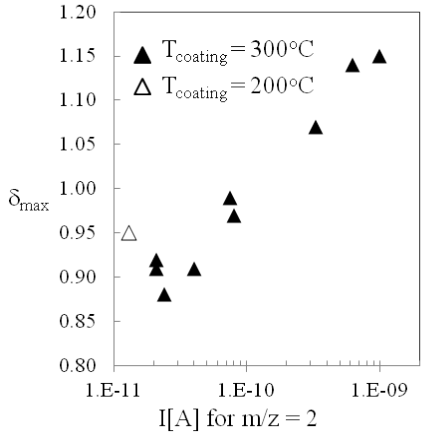

Figure 9: $\delta_{max}$ in function of the residual gas analyser signal for hydrogen for coatings done by DCCMS. (From [6])

The evolution of $\delta_{max}$ with the time of air exposure, often called ageing, was studied for films issued from the different coating techniques and stored in different conditions: in air inside a polystyrene box; in air and wrapped in aluminium foil inside a polystyrene box and in static vacuum inside a stainless steel chamber. Samples directly stored in polystyrene box aged considerably while samples wrapped in aluminium foil have negligible ageing (Figure 10). The ageing mechanism is not yet understood, but the coating can be effectively protected by simple means (aluminium foil, flanges, etc).

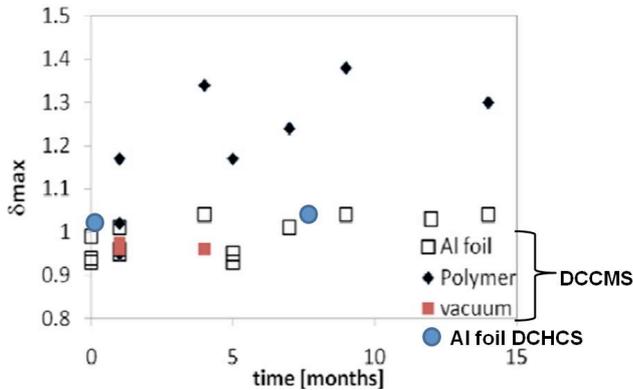

Figure 10: Evolution of $\delta_{max}$ in function of the time of air exposure. (From [7], updated for DCHCS coatings.)

### Electron Multipacting induced by a RF standing wave in laboratory (Multipactor)

The effectiveness of the real size carbon coatings can be assessed by comparing the amount of EM induced by an RF standing wave in coated and uncoated dipoles. The RF is injected via a tungsten wire stretched inside the beampipe and the reflected power is measured as a function of the input power. In case of EM, the system goes out of tune and the reflected power increases abruptly. A pressure burst is also observed and the composition of the gas released, (hydrogen, carbon monoxide and carbon dioxide), correspond to electron stimulated desorption, confirming the EM. Details of the setup and measurements can be found in [4]. Up to now the two MBB dipoles tested shows a dramatic decrease in EM after applying the coating.

### Electron Cloud Monitors in the SPS

EM activity in the SPS is measured in electron cloud monitors. Details of the setup can be found in [8]. The ECM signals measured in the coated liners are about four orders of magnitude below the one in the stainless steel reference, confirming the efficiency of the carbon coating [5]. Some of these liners have remained up to three years in the SPS, and have been vented to air a few times during technical stops and winter shutdowns. They went through several machine development (MD) runs without any measurable degradation in the signal of the ECMs. This is in agreement with the values of the SEY measured in samples cut from the liners extracted from the SPS: the ageing is negligible and very close to the accuracy of the SEY measurement (+/- 0.03), (Table 1).

Table 1. Some of the carbon coated liners tested in the SPS. $\delta_{max}$ *initial* was measured from a witness sample a few hours after the coating was vented to air; $\delta_{max}$ *extracted* was measured in a sample cut from the liner a few days after removal from the SPS.

| Liner type | Time in SPS | $\delta_{max}$ initial | $\delta_{max}$ extracted |
|---|---|---|---|
| Stainless steel (reference) | 1 year (5MD runs) | 2.25 | 1.72 |
| Carbon stripe 40 mm wide | 1 year (5MD runs) | 0.92 | 0.97 |
| Carbon on top of Zr | 1.5 years (9 MD runs) | 0.95 | 0.99 |
| Carbon (CNe64) | 3 months (2 MD runs) | 0.95 | 0.97 |
| Carbon (CNe65) | 3 months (2 MD runs) | 0.95 | 0.97 |

## Dynamic pressure measurements in the SPS

Another way to check the effectiveness of the coating against EM in the SPS is to monitor the time dependence of the pressure in the presence of beam. To do such experiments a drift section of the SPS was coated with carbon. At each extremity was added a short stainless steel tube (0.3 meter on one side and 0.8 meter on the other side). Independent solenoids were wrapped around the three sections. When the solenoids are powered on the secondary electrons generated at the walls of the beampipes are trapped and cannot contribute to EM built up. Three vacuum gauges, (penning discharge type), monitor the pressure: gauge 1 in front of the first stainless steel section; gauge 2 in the middle of the 12 meter carbon coated section; gauge 3 in front of the second stainless steel section. Two ion pumps, (Leybold IZ series, 25 l/s), are placed at both extremities of the setup. A schematic view and photos are shown in Figure 11.

When all the solenoids are off, the gauges register a peak pressure of about $3.6 \times 10^{-7}$ mbar, (Figure 12), for each cycle of the beam (injection, acceleration, dump). From the moment the solenoids on the stainless steel sections are powered on, the EM in these sections is suppressed and the maximal peak pressures decreases to about $2.6 \times 10^{-7}$ mbar.

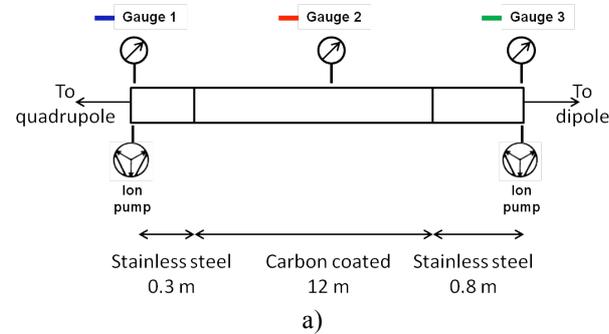

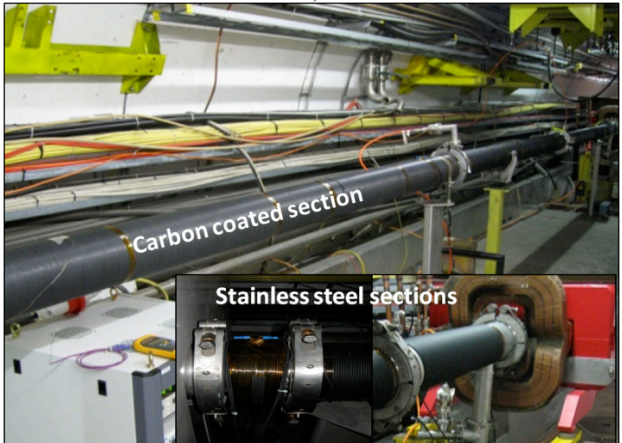

Figure 11: View of the setup in the SPS a) schematic; b) photos.

No further decrease is observed by switching on the solenoid on the carbon coated tube. This proves that there is no EM to be suppressed. Keeping the solenoid on the carbon coated tube on and switching off the solenoids on the stainless steel parts results in the peak pressure returning to its initial value of about $3.6 \times 10^{-7}$ mbar, confirming that the solenoid on the carbon coated tube does not have any influence on the pressure rise and, consequently, the observed EM is not in this tube. The slight slope on the pressure observed after powering the carbon coated solenoid is due to thermal degassing induced by the heating of the coil.

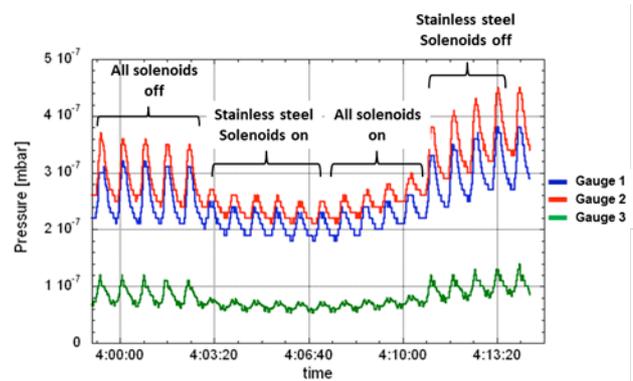

Figure 12: Dynamic pressure variation for the different combinations

## DISCUSSION AND CONCLUSIONS

As seen from Figure 1, all the coatings obtained by sputtering have a $\delta_{max} \sim 1.0$ while those obtained by PECVD have $\delta_{max} \sim 1.5$. We believe this difference is related to the amount of hydrogen in the coating. Hydrogen favours the formation of $sp^3$ bonds, [9], resulting in Diamond Like Coatings (DLC) and these types of coatings are known for having high SEY [10]. During deposition by PECVD, the $C_2H_2$ molecule is dissociated by the electrons in the plasma into $C_2H$ and H [11]. The $C_2H$ radical is the main growth precursor of the film and the coatings produced by this technique are intrinsically charged with hydrogen. In addition, the available hydrogen can also react with dangling bonds. For this reason, the PECVD technique was discarded for the production of carbon coatings with the aim of achieving low SEY.

In the case of sputtered coatings, the partial pressure of hydrogen available during the growth is orders of magnitude lower, resulting in films with a higher fraction of $sp^2$ type bonds, (typical of graphite that has low SEY), and with $\delta_{max} \sim 1.0$. This analysis is corroborated by the correlation between the $\delta_{max}$ of the sputtered films and the partial pressure of hydrogen during the coating (Figure 9). For this reason some sputtering setups may require an efficient pumping of hydrogen during film growth in order to reduce its partial pressure. Such is the case for the coating of the MBA beampipes by DCCMS, where the co-deposition of a getter, (Figure 3), assures a distributed pumping speed.

The reason why the SEY of sputtered films is lower than that of pure graphite (HOPG) is not yet understood (see Figure 1). One hypothesis is that the higher roughness of the films relative to HOPG, (see Figure 7), lowers the effective SEY.

So far, sputtered coatings passed all tests successfully: direct measurements of EM in the SPS, (using the ECM strip detectors); dynamic pressure with LHC type beam; vacuum pumpdown and EM induced by RF. Ageing is also acceptable. In air it is negligible if the coatings are protected by aluminium foil or flanges ($\delta_{max}$ < 1.15 after more than one year of air exposure). Storage in nitrogen atmosphere or desiccators also prevents ageing. In the SPS, samples cut from liners exposed to the beam showed almost no signs of ageing, ($\delta_{max}$ < 1.0 after more than one year in the machine), in agreement with the results obtained with ECMs: no increase in the EM signal along the time.

The techniques based on planar magnetron, (DCPMS), were discarded because the coatings were not satisfactorily uniform. The one using the dipole's field due to the leak of electrons at the cathodes extremities; the one using arrays of permanent magnets because the glow discharges were too sensitive to minor variations of the distance between the cathode surface and the magnetic circuits. Instead, the hollow cathode setup, DCHCS, exhibits very stable and reproducible plasma, resulting in a uniform coating. Two MBB dipoles have already been coated without disassembling / assembling. One is installed in the SPS and the other one was successfully tested in the *Multipactor* system. This technique is almost mature for large scale production. In 2013, during the first long shutdown of the LHC, 14 dipoles must be coated and installed in the machine in order to have two full SPS cells coated with carbon. Other parts, like pumping port shields and beampipes for the quadrupoles will be coated by DCCMS. The goal is to run the SPS from 2014 to 2017 with two coated cells. This will be the ultimate test for the carbon coatings before considering the technology ready to coat the entire SPS.